\documentclass[
    ,final            % use final for the camera ready runs
  ]
  {aipproc}

\layoutstyle{8x11double}

\def\bfbeta{\beta \kern-.680em \beta}

\def\bfomega{\omega \kern-.60em \omega}

\def\bfpsi{\psi \kern-.645em \psi}

\def\bfrho{\rho \kern-.55em \rho}

\def\bftau{\tau \kern-.524em \tau}

\def\bfzeta{\zeta \kern-.50em \zeta}

\def\bfkappa{\kappa \kern-.50em \kappa}

\def\bfdel{\nabla \kern-.770em \nabla}
\begin{document}

\title{Turbulence, Energy Transfers and Reconnection in Compressible 
Coronal Heating Field-line Tangling Models}


\author{R. B. Dahlburg}{
  address={Code 6440, Naval Research Laboratory, Washington, DC 20375 USA}
}

\author{A. F. Rappazzo}{
  address={Jet Propulsion Laboratory, California Institute of Technology,  Pasadena, CA 91109 USA}
}

\author{M. Velli}{
  address={Jet Propulsion Laboratory, California Institute of Technology,  Pasadena, CA 91109 USA}
}

\begin{abstract}
MHD turbulence has long been proposed as a  mechanism for the heating of coronal loops in the 
framework of the Parker scenario for coronal heating. So far most of the studies have focused 
on its dynamical properties without considering its thermodynamical and radiative features,
because of the very demanding computational requirements.
In this paper we extend this previous research to the compressible regime, including an energy
equation, by using HYPERION, a new parallelized, viscoresistive, three-dimensional 
compressible MHD code. HYPERION employs a Fourier collocation -- finite difference 
spatial discretization, and uses a third-order Runge-Kutta temporal discretization.   
We show that the implementation of a thermal conduction parallel to the DC magnetic
field induces a radiative emission concentrated at the boundaries, with properties
similar to the chromosphere--transition region--corona system. 
\end{abstract}
\maketitle

%%%%%%%%%%%%%%%%%%%%%%%%%%%%%%%%%%%%%%%%%%%%
%% MAINMATTER
%%%%%%%%%%%%%%%%%%%%%%%%%%%%%%%%%%%%%%%%%%%%

\section{INTRODUCTION}
\vspace{-10pt}

Magnetohydrodynamic turbulence in the framework of the Parker scenario for coronal 
heating \cite{p72,p88} has been a very challenging problem to
investigate  numerically \cite{dm99,ev99,ev96,hvh96,rved07,rved08}.
As computers have advanced, it has become more feasible to do the large storage compressible problem.  

Why is it important to include compressibility and its related effects?  There are three basic categories of 
interest with respect to active region loops and the coronal heating problem, \emph{viz.}: 
structural, dynamical and 
thermodynamical.  The most significant structural effect is stratification due to gravity.  We can also modify this 
term to model the curvature of a typical loop.   Among new dynamical effects that are possible are compression 
and rarefaction of the plasma, as well as the formation of shocks.  

Thermodynamical effects include thermal 
conduction and radiation.  In addition, the diffusivities can be temperature dependent.   It's important to have 
these features in the model to begin to reproduce the energy cycle:  kinetic energy in the photosphere is 
transformed into magnetic energy in the corona by means of photospheric footpoint convection. It is then
transported to small scale by MHD turbulence, where through magnetic reconnection it is converted
into thermal, kinetic and perturbed magnetic energies.  
Heat is then conducted from the high temperature corona back toward the low 
temperature photosphere, where it is lost \emph{via} optically thin radiation.

Incompressible and cold plasma models only contain the first few parts of this energy cycle, without taking
into account the thermodynamics.
%i.e.\ fluctuating magnetic energy is induced through footpoint motions and energy transport and conversion by 
%means of magnetic reconnection.  
Any magnetic energy lost through Ohmic diffusion and any kinetic energy lost through viscous 
diffusion is simply lost from the system and the physics involved with thermal conduction and radiation is 
irrelevant.

Our new compressible code HYPERION has allowed us to make a start at examining the fully compressible, 
three-dimensional Parker coronal heating model.  HYPERION is a parallelized Fourier collocation--finite 
difference code with third-order Runge-Kutta time discretization that solves the compressible MHD equations 
with DC field--aligned thermal conduction and radiation included.  

\vspace{-10pt}
\section{SETTING UP THE PROBLEM}

\vspace{-10pt}
\subsection{Governing equations}
\vspace{-10pt}

We model the solar corona as a compressible, dissipative magnetofluid. The equations which govern such a system, written here in a dimensionless
form, are:{\setlength\arraycolsep{-2.5pt}
\begin{eqnarray}
&&\frac{\partial \rho}{\partial t}=-\nabla\cdot (\rho {\bf v})\\[.4em]
&&\frac{\partial \rho {\bf v}}{\partial t}=-\nabla\cdot({\rho\bf v v}) 
   -{\beta}\nabla p + {\bf J}\times{\bf B}+
 \frac{1}{S_v}\nabla\cdot\mathbf{\zeta}\\
&&\qquad \qquad \qquad \qquad \qquad \qquad  \qquad  + \rho g(z) {\bf\hat e}_z\\[.4em]
&&\frac{\partial T}{\partial t}= -{\bf v}\cdot\nabla T  
 - (\gamma - 1) (\nabla\cdot {\bf v}) T
+\frac{1}{ Pr\, S_v }\frac{1}{\rho}\frac{\partial^2 T}{\partial z^2} + f\\[.4em]
&&\frac{\partial {\bf B}}{\partial t}=\nabla\times{\bf v}\times{\bf B} 
  + \frac{1}{S}\nabla\times \nabla\times {\bf B} \\[.4em]
&& \nabla\cdot{\bf B} = 0.
\end{eqnarray}
}where
$f={(\gamma -1)\over\beta\rho} [
{ 1\over S_v} \zeta_{ij}{\bf e}_{ij}
+{1\over S} (\nabla\times{\bf B})^2 
-{1\over S_r} \rho^2\Lambda (T) ].$
The system is closed by the equation of state, 
\begin{equation}
p = \rho T.
\end{equation}   %$$ p = \rho T .\eqno(6)$$ 
In the preceding equations the variables are defined in the following way:~~
$\rho ({\bf x}, t)$ is the mass density,
${\bf v}({\bf x}, t) = (u, v, w)$ is the flow velocity,
$p({\bf x}, t)$ is the thermal pressure,
${\bf A}({\bf x}, t) = (A_x, A_y, A_z)$ is the magnetic vector potential,
${\bf B}({\bf x}, t) = (B_x, B_y, B_z) = \nabla\times{\bf A}$ is the magnetic induction field
expressed in terms of the associated Alfv\'en velocity (${\bf B} \rightarrow {\bf B}/\sqrt{4\pi\rho_0}$),
${\bf J} = \nabla\times{\bf B}$ is the electric current density,
$T({\bf x}, t)$ is the plasma temperature,
$\zeta_{ij}= \mu (\partial_j v_i + \partial_i v_j) - 
\lambda \nabla\cdot {\bf v} \delta_{ij}$ is the viscous stress tensor,
$e_{ij}= (\partial_j v_i + \partial_i v_j)$ is the strain tensor,
and $\gamma$ is the adiabatic ratio.
The thermal conductivity ($\kappa$), magnetic resistivity $(\eta)$, and shear viscosity $(\mu)$
are assumed to be constant and uniform, and Stokes relationship is assumed  
%\ref{R. B. Dahlburg and J. M. Picone, {\sl Phys. Fluids B} {\bf 1}, 2153 (1989).|DP|}, 
so the bulk viscosity $\lambda = (2/3) \mu$.
The function $g(z)$ defines the gravitational field strength: 
at $t=0$ we define $\rho$ as
$\rho_0  \exp[-{1\over 2} g \cos({\pi z\over L_z})]$.
Assuming a uniform temperature we can determine the gravity as
$g = \beta {1\over\rho}{d\rho\over dz}.$
The function $\Lambda(T)$ describes the temperature dependence of the radiation
($\Lambda(T) =0$ for $T<T_0$):
\begin{equation}
\Lambda(T) =  \frac{T-T_0}{T_0} e^{(\epsilon T-T_0)/(\tau T_0)}, \qquad T \ge T_0
\end{equation}
where $T_0$ is the wall (photospheric) temperature, 
$\epsilon=2$ and $\tau=2$ (\cite{ddpmk87,dm88}).
The important dimensionless numbers are:
$S_v = \rho_0 V_A L_0 / \mu \equiv$ viscous Lundquist number, 
$S = V_A L_0 / \eta \equiv$ Lundquist number,
$S_r = V_A L_0 / \chi \equiv$ radiative Lundquist number (the new parameter $\chi$ determines the strength of the radiation),
$\beta = p_0 / B_0^2 \equiv$ pressure ratio at the wall,
$Pr = C_p \mu / \kappa \equiv$ Prandtl number, and
$A = V_A / V_0 \equiv$ Alfv{\'e}n number.
In these definitions, $\rho_0$ is a characteristic density, $V_A$ is the vertical Alfv{\'e}n speed
(\emph{used as the characteristic velocity to render velocities dimensionless}), 
$L_0$ is the vertical box length ($=L_z$), $C_p$ is
the specific heat  at constant pressure, $C_s$ is the free-stream sound speed, and $V_0$ is
the characteristic flow speed. Time ($t$) is measured in units of Alfv\'en transit times
($=L_0/V_A$).

\vspace{-10pt}
\subsection*{Boundary Conditions and Forcing}
\vspace{-10pt}

We solve the governing equations in a box of dimensions ($L_x, L_y, L_z$).
The system has periodic boundary conditions in $x$ and $y$, and line-tied boundary conditions in $z$.
To model a section of a coronal loop the system is threaded by a DC magnetic field in
the $z$-direction ($B_0$).  

We then employ a simple, three-dimensional extension of
the time-dependent forcing function used in the previous studies \cite{hvhms96,ev99}, {\it i.e.,} at the top and bottom walls we evolve a stream function:
\begin{equation}
\psi_{nm} (x, y, t) = f_1 \sin^2 \left(\frac{\pi t}{2 t^*}\right) + f_2  \sin^2 \left(\frac{\pi t}{2 t^*} + \frac{\pi}{2}\right)
\end{equation}
where
$f_i (x,y) = V_0 \sum_n\sum_m a_{nm}^i sin(k_n x + k_m y + \zeta_{nm}^i).$
 \noindent
Values for $k$ are given by $3\le(k_n^2 + k_m^2)^{1\over 2}\le4$.
At the top and bottom walls the magnetic vector potential is convected by the resulting flows.

%We then employ a generic forcing function similar to those used in the %previous studies of incompressible systems \cite{ev99}, {\it i.e.,} at the top and %bottom walls we define a time-independent stream function:
%$$\psi_{nm} (x,~y,~t) = \sum_n\sum_m a_{nm} \sin(k_n x + k_m y + 
%\zeta_{nm})
% \eqno(7)$$
 %\noindent
 %Here $a_{nm}$ are random amplitudes and $\zeta_{nm}$ are random %phases.  %We use the same values for these amplitudes and phases in both %the 3DMHD code and the RMHD code.
%Values for $k$ are given by $3\le(k_n^2 + k_m^2)^{1\over 2}\le4$.

That is, the line-tied boundary conditions are:
$$\rho(\pm L_z/2) = \rho_0,  $$
$$\rho u(\pm L_z/2) = - \rho_0\partial\psi / \partial y,  $$
$$\rho v(\pm L_z/2) = \rho_0\partial\psi / \partial x,  $$
$$\rho w(\pm L_z/2) = 0, $$
$$\partial A_x / \partial t |_{\pm L_z/2}= v~B_0, $$
$$\partial A_y / \partial t |_{\pm L_z/2} = -u~B_0, $$
$$B_z(\pm L_z/2) = B_0, $$
$$T(\pm L_z/2) = T_0 . $$
\medskip\noindent
The enforcement of the boundary conditions is discussed in greater detail in \cite{dlkn07}.  

%Note that the use of a time-dependent wall stream function is possible:
%NORMALIZATION??
\vspace{-10pt}
\subsection*{Numerics}
\vspace{-10pt}
Equations 5 and 6 can be replaced by the magnetic vector potential equation:
\begin{equation}
\frac{\partial{\bf A}}{\partial t} = {\bf v}\times\nabla\times{\bf A} 
  + \frac{1}{S}~ \nabla\times\nabla\times {\bf A}
\end{equation}
where ${\bf A} = \nabla\times {\bf B}.$ 
%\cite{dan97},\cite{lda02}(Dahlburg, Antiochos and Norton 1997; Linton, Dahlburg and Antiochos 2002). 
Thus we solve numerically the equations 1-3 and 7 together with equation 6.  Space is discretized in $x$ and 
$y$ with a Fourier collocation scheme  \cite{dp89} with isotropic truncation dealiasing.  
Spatial derivatives are calculated in the appropriate transform space, and nonlinear product terms are 
advanced in configuration space.  A second-order central difference technique \cite{dmz86} is used for the 
discretization in $z$.  
A staggered mesh also is employed in the $z$-direction\cite{s87}.   In general, the fields that are defined at the 
$z$ boundaries are advanced in time on the standard mesh.  Other quantities of interest are defined and 
advanced in time on the staggered mesh.
That is, on the standard mesh we look at $\rho, \rho u, ~\rho v, ~\rho w, ~A_x, ~A_y,~B_z$ and $T$.  Some 
derived fields such as $\omega_x, ~\omega_y, ~\omega_z,  ~j_x,$ and $j_y$ are also defined on the standard 
mesh.  On the staggered mesh we look at $A_x, ~B_x,~ B_y,$ and $j_z$.  Note that for plotting purposes we 
interpolate these latter fields onto the standard mesh (at the boundaries an extrapolation is performed).

A time-step splitting scheme is employed.  All terms, with the exception of the vertical pressure gradient and the 
gravitation term, are discretized in time with a third-order Runge-Kutta scheme.  
The pressure step for the $z$-momentum is solved with a second-order Lax-Wendroff one-step 
central difference scheme.   The vertical gravitation term is advanced using the forward Euler method.
 
The code has been parallelized using MPI.  A domain decomposition is employed in which the computational 
box is sliced up into $x$--$y$ planes along the $z$ direction.

%%%%%%%%%%%%%%%%%%%%%%%%%%%%%%%%%%%%%%%%%%%%
%% SAMPLE TABLE
%%
%% Shows the use of \tablehead and \tablenote
%% macros
%%%%%%%%%%%%%%%%%%%%%%%%%%%%%%%%%%%%%%%%%%%%

%\begin{table}
%\begin{tabular}{lrrrr}
%\hline
%  & \tablehead{1}{r}{b}{Single\\outlet}
%  & \tablehead{1}{r}{b}{Small\tablenote{2-9 retail outlets}\\multiple}
%  & \tablehead{1}{r}{b}{Large\\multiple}
%  & \tablehead{1}{r}{b}{Total}   \\
%\hline
%1982 & 98 & 129 & 620    & 847\\
%1987 & 138 & 176 & 1000  & 1314\\
%1991 & 173 & 248 & 1230  & 1651\\
%1998\tablenote{predicted} & 200 & 300 & 1500  & 2000\\
%\hline
%\end{tabular}
%\caption{Average turnover per shop: by type
%  of retail organisation}
%\label{tab:a}
%\end{table}

\begin{figure}
  \includegraphics[height=.30\textheight]{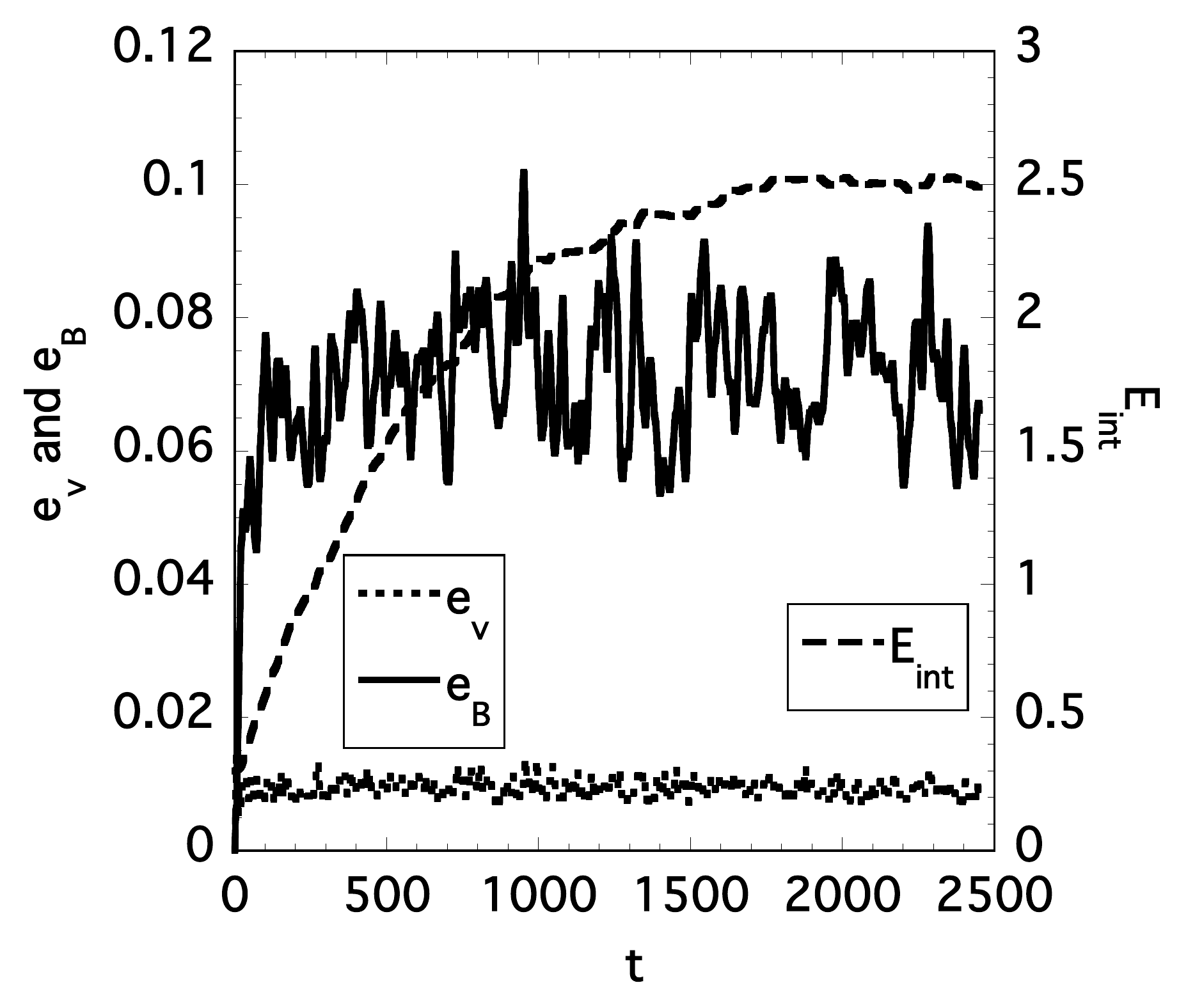}
  \caption{Energies vs.\ time. Time is
measured in units of axial Alfv\'en crossing times $L_z/V_A$.}
\end{figure}

\vspace{-10pt}
\section{New results}
\vspace{-10pt}
In this section we report on the results of a preliminary numerical simulation 
of the model.    This simulation is run with $L_x = 2\pi, L_y=2\pi$ and $L_z = 8\pi$.  
Other important parameters are $g=6.0, \gamma=5/3, S = S_v = 80000, \beta=0.001, B_0=1.0, V_0 = 0.01 
(\sqrt{2}/ 2), A = V_A/V_0 = 100\sqrt{2}, \rho_0=1.0, T_0=1.0, t^*=20.0, Pr = 0.001$, and $S_{r} = 0.0004$.

\vspace{-10pt}
\subsection{Temporal diagnostics}
\vspace{-10pt}
To insure that we are obtaining good statistics, the system has to settle down into a steady state.  Evidence for 
this is shown in Figure~1, which shows some of the important energies as functions of time (time is
expressed in units of Alfv\'en transit times $L_0/V_A$).   As  seen in our 
previous RMHD simulations, the fluctuating magnetic and kinetic energies 
($e_v ={1\over 2}\int_{-L_z / 2}^{L_z / 2}\int_0^{L_y}\int_0^{L_x} |{\bf v}|^2~dx~dy~dz$
and
$e_b={1\over 2}\int_{-L_z / 2}^{L_z / 2}\int_0^{L_y}\int_0^{L_x} |{\bf b}|^2~dx~dy~dz$)
settle down pretty quickly, with $e_b > e_v$ (these quantities are also time intermittent).  
Note, however, that the total internal energy 
$E_{int}={\beta\over \gamma - 1}\int_{-L_z / 2}^{L_z / 2}\int_0^{L_y}\int_0^{L_x} \rho~T~dx~dy~dz$
takes much longer to level off in time.  
This reflects the fact that the system must heat up to attain the driven-dissipative steady state.

The quantities shown in Figure~2 provide temporal information about the dissipation.  
Note that the radiation loss is a new quantity respect to previous simulations.
Shown are the enstrophy
$\Omega={1\over S_v}\int_{-L_z / 2}^{L_z / 2}\int_0^{L_y}\int_0^{L_x} |{\bf \omega}|^2~dx~dy~dz$,
mean square electric current
$J=\frac{1}{S}\int_{-L_z / 2}^{L_z / 2}\int_0^{L_y}\int_0^{L_x} |{\bf j}|^2~dx~dy~dz$,
and the total radiation losses
$D={1\over S_{r}}\int_{-L_z / 2}^{L_z / 2}\int_0^{L_y}\int_0^{L_x} \rho^2~\Lambda (T)~dx~dy~dz$
as functions of time. The first two quantities behaves similarly to previous RMHD simulations,
while the radiative losses settles on a similar timescale than the internal energy (Figure~1).
\begin{figure}
  \includegraphics[height=.30\textheight]{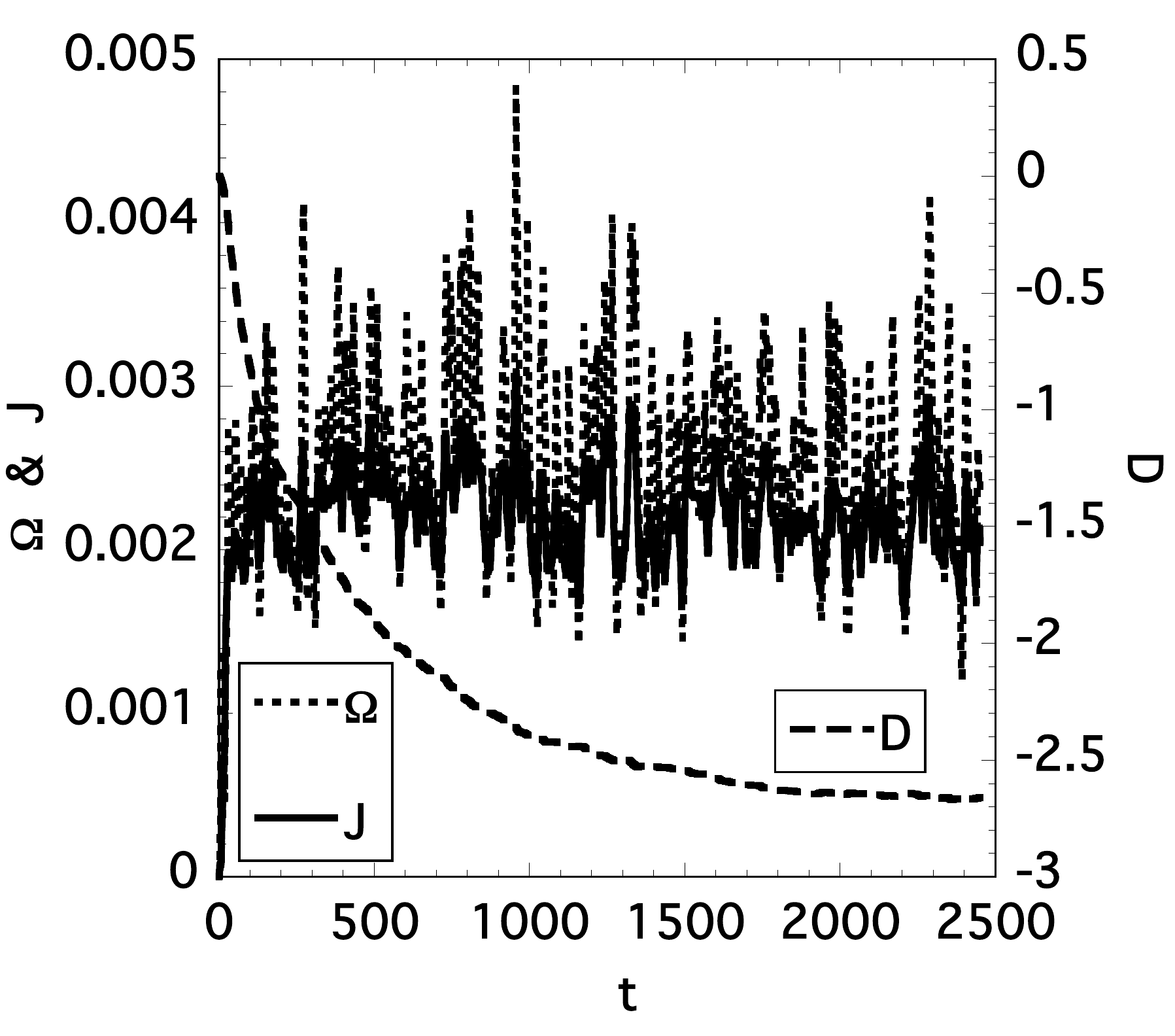}
  \caption{Dissipation vs. time. Time is
measured in units of axial Alfv\'en crossing times $L_z/V_A$.}
\end{figure}
 \vspace{-10pt}
\subsection{Spatial diagnostics}
\vspace{-10pt}
The following  quantities are averaged over the perpendicular directions ($x$ and $y$) and also over $1000\le t \le2000$.  We look at these to determine the times averaged state of the system under unsteady heating.

We first look at some of the quantities related to the dissipation of the system.
Figure~3 shows some of the time averaged quantities as a function of $z$, the direction of the large magnetic field $B_0$.
Shown are the time averaged dissipation intensity for the parallel vorticity
$Q_z (z)=< [ \int_0^{L_y}\int_0^{L_x} {\omega_z^2} (x, y, z)\, dx dy ]^{1\over 2}>$
and also for the parallel electric current
$G_z (z)=<[\int_0^{L_y}\int_0^{L_x} {j_z}^2 (x, y, z)\, dx dy]^{1\over 2}~>$
as well as the time averaged mean radiation rate
$D_m (z)={1\over S_{r}}< \int_0^{L_y}\int_0^{L_x}\rho^2\Lambda (T)  (x, y, z)\, dx dy >$.
As in previous simulations, current density and vorticity are aligned to the dc magnetic field,
so that their $z$-components are strongly dominant.
Note that ``$<~~>$'' denotes the time averaging.
The symmetry in $z$ of these quantities indicates that we have averaged over a sufficient period of time. 
\begin{figure}
  \includegraphics[height=.26\textheight]{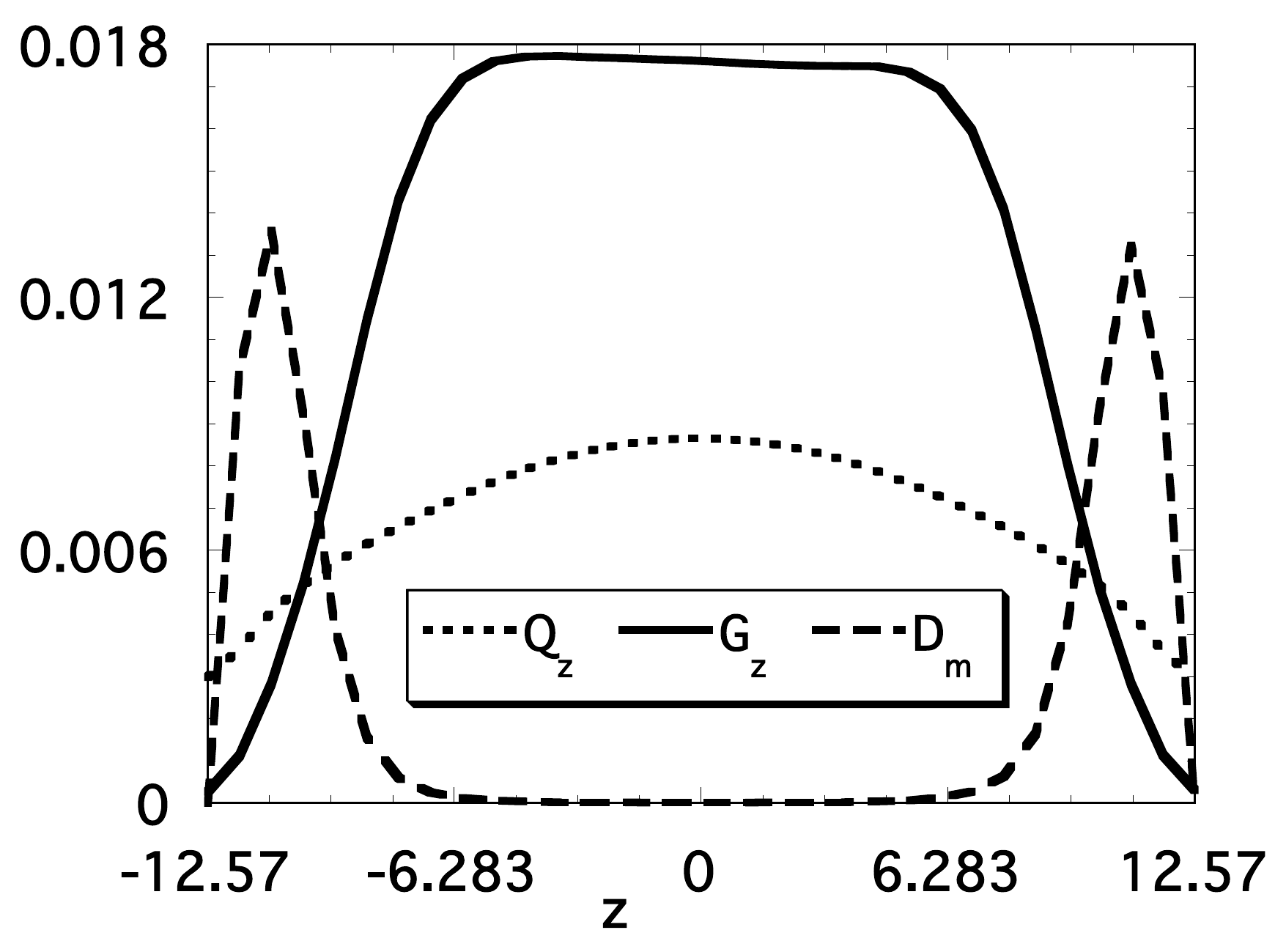}
  \caption{Time averaged dissipation rates (viscous $Q_z$, ohmic $G_z$ and radiative $D_m$)
    as a function of the axial coordinate z.}
\end{figure}

In Figure~4 we take a look at the time averaged thermodynamic state for the 
unsteady heating case:  shown are the
time averaged mean mass density
$\rho_m(z)=< \int_0^{L_y}\int_0^{L_x}\rho (x, y, z) dx dy >$
and the time averaged mean temperature
$T_m(z)=< \int_0^{L_y}\int_0^{L_x}T (x, y, z) dx dy >$
as functions of $z$.

The density profile is a result of the gravitational density stratification.
Figures~3 and 4 show feature typical of the chromosphere-transition region-corona
system, where density increases at lower heights, while temperature increases in the
high corona. Notice that most of the ohmic and viscous dissipation ($Q_z$ and $G_z$) 
takes place in the high corona, while radiation ($D_m$) origins mostly near the boundaries,
where it is peaked (see Figure~3). This mostly results from the higher density values near
the boundaries, as $\Lambda(T)$ is multiplied by $\rho^2$ in the radiative term $D_m$.

\vspace{-10pt}
\section{DISCUSSION}
In this paper we have presented some preliminary results of our simulations of compressible DC coronal heating using our new HYPERION code. The inclusion of a thermal conductivity parallel to the
DC magnetic field, coupled with a gravitational density stratification, gives rise to temperature
and radiation features typical of a realistic coronal loop. 

This is an encouraging starting point to investigate the thermodynamical properties of a coronal loop
threaded by a strong magnetic field whose footpoints are shuffled by photospheric motions.

\begin{figure}
  \includegraphics[height=.28\textheight]{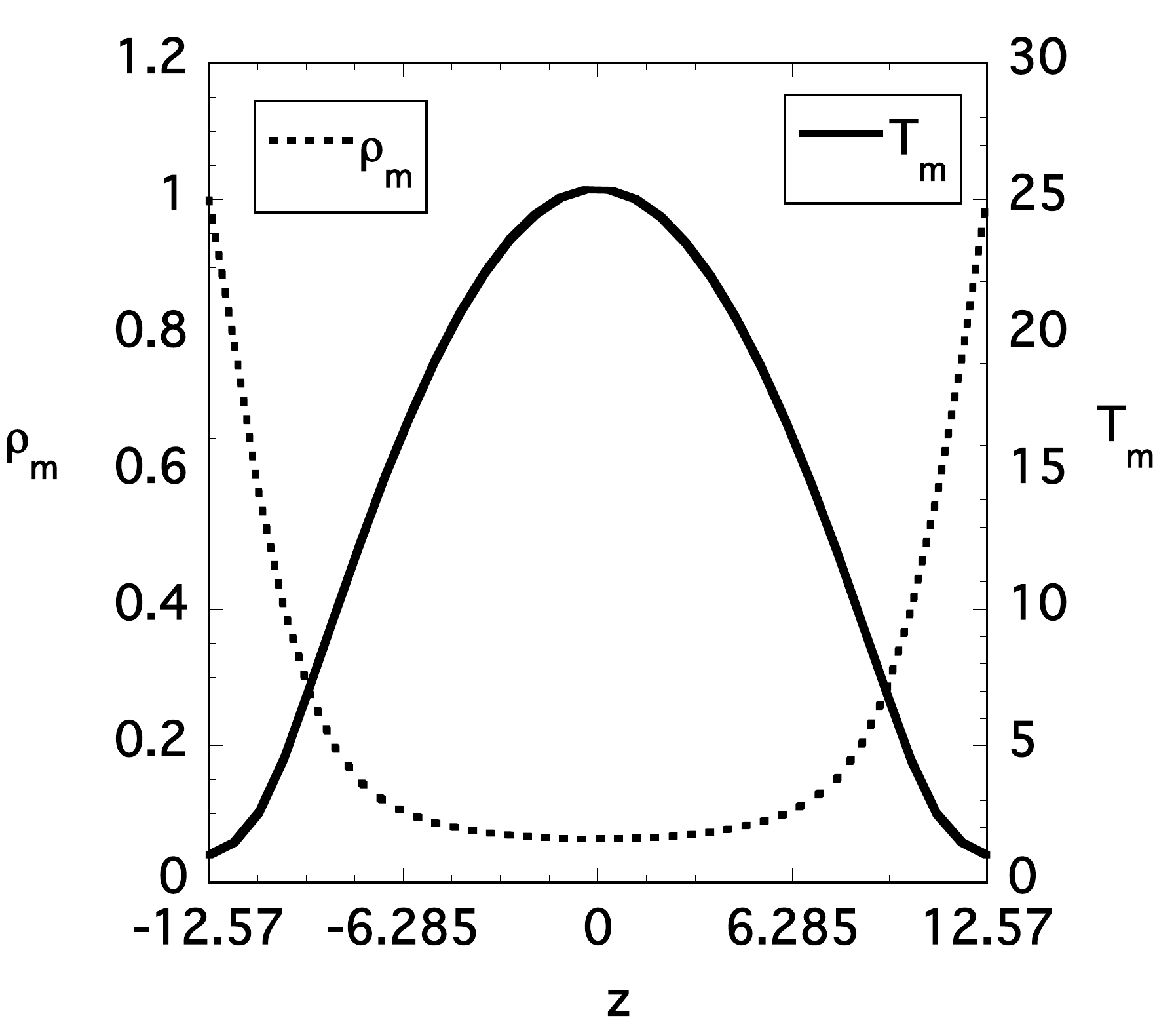}
  \caption{Time averaged mass density ($\rho_m$) and temperature ($T_m$) as a function of z.}
\end{figure}

%%%%%%%%%%%%%%%%%%%%%%%%%%%%%%%%%%%%%%%%%%%%%%%%
%% BACKMATTER
%%%%%%%%%%%%%%%%%%%%%%%%%%%%%%%%%%%%%%%%%%%%%%%%

\vspace{-0pt}
\begin{theacknowledgments}
We thank J. A. Klimchuk, G. Einaudi, G. Nigro and H, Warren for helpful conversations. 
This work was supported by ONR and the NASA Sun-Earth Connection Theory and Guest Investigator 
Programs.  A.F.R. is supported by the NASA postdoctoral program.
Computer time was provided by the Department of Defense High Performance Computing Modernization 
Program. The research described in this paper was carried out in part at the Jet Propulsion Laboratory, 
California Institute of Technology, under a contract with the National Aeronautics and Space Administration. It 
was also supported by the Italian Space Agency contract Solar System Exploration.

\end{theacknowledgments}
 \vspace{-0pt}

\bibliographystyle{aipproc}

\end{document}